\documentstyle[12pt]{article}
\newcommand{\nc}{\newcommand}
\nc{\drm}{\rm d} \nc{\om}{\omega} \nc{\bb}{\begin{equation}} \nc{\ee}{\end{equation}}
\nc{\bega}{\begin{eqnarray}} \nc{\ega}{\end{eqnarray}}
\nc{\begae}{\begin{eqnarray*}} \nc{\egae}{\end{eqnarray*}} \nc{\ga}{\gamma}
\nc{\ia}{{\bf i}} \nc{\age}{\dagger} \nc{\sig}{\sigma} \nc{\Sig}{\Sigma}
\nc{\longr}{\longrightarrow} \nc{\al}{\alpha} \nc{\vare}{\varepsilon}
\nc{\C}{I$\!\!\!$C} \nc{\la}{\lambda} \nc{\1}{1\!\!1} \nc{\lan}{\langle}
\nc{\de}{\delta} \nc{\ran}{\rangle} \nc{\h}{\hspace*{0.5 cm}}
\nc{\hs}{\hspace*} \nc{\dis}{\displaystyle} \nc{\wid}{\widetilde}
\nc{\wide}{\widehat} \nc{\tm}{\times} \nc{\pa}{\partial} \nc{\sta}{\stackrel}
\nc{\cent}{\centerline} \nc{\vs}{\vspace*} \nc{\Lef}{\Leftrightarrow}
\nc{\wede}{\wedge} \nc{\Vbf}{\mbox{\boldmath $V$}} \nc{\Abf}{\mbox{\boldmath $A$}} \nc{\Ebf}{\mbox{\boldmath $E$}}
\nc{\Hbf}{\mbox{\boldmath $H$}} \nc{\Mbf}{\mbox{\boldmath $M$}} \nc{\Pbf}{\mbox{\boldmath $P$}}
\nc{\sbf}{\mbox{\boldmath $s$}} \nc{\pabf}{\mbox{\boldmath $\partial$}} \nc{\del}{\mbox{\boldmath $\delta$}}
\nc{\gabf}{\mbox{\boldmath $\gamma$}} \nc{\etbf}{\mbox{\boldmath $\eta$}}
\nc{\mubf}{\mbox{\boldmath $\mu$}} \nc{\ombf}{\mbox{\boldmath $\om$}}
\nc{\Sigbf}{\mbox{\boldmath $\Sigma$}} \nc{\vbf}{\mbox{\boldmath $v$}}
\nc{\abf}{\mbox{\boldmath $a$}} \nc{\albf}{\mbox{\boldmath $\alpha$}}
\nc{\wbf}{\mbox{\boldmath $w$}} \nc{\xbf}{\mbox{\boldmath $x$}}
\nc{\ubf}{\mbox{\boldmath $u$}} \nc{\rbf}{\mbox{\boldmath $r$}}
\nc{\jbf}{\mbox{\boldmath $j$}} \nc{\imp}{\mbox{\boldmath $p$}}
\nc{\z}{\`} \nc{\dopsi}{\dot{\psi}}
\nc{\dox}{\dot{x}} \nc{\dopi}{\dot{\pi}} \nc{\dop}{\dot{p}}
\nc{\dov}{\dot{v}} \nc{\qq}{\qquad \qquad} \nc{\ii}{\`{\i \ }} \nc{\iv}{\"{\i}}
\nc{\psip}{\psi^{\dagger}} \nc{\psis}{\psi^{\star}} \nc{\G}{{\wide G}} \nc{\A}{{\wide A}}
\nc{\Ho}{{\wide H}} \nc{\vo}{{\wide {\vbf}}} \nc{\po}{{\wide {p}}} \nc{\jo}{{\wide {\jbf}}}
\nc{\xo}{{\wide {\xbf}}} \nc{\impo}{\wide {\imp}} \nc{\hh}{{\hbar\over 2}} \nc{\xibf}{\mbox{\boldmath $\xi$}}
\nc{\Xbf}{\mbox{\boldmath $X$}} \nc{\Rbf}{\mbox{\boldmath $R$}} \nc{\ug}{\; = \;}
\nc{\erm}{{\rm e}} \nc{\bi}{\bibitem} \nc{\bt}{\beta} \nc{\ov}{\over}
\nc{\ddopsi}{\ddot{\psi}} \nc{\dopsib}{\dot{\overline{\psi}}}
\nc{\ddopsib}{\ddot{\overline{\psi}}} \nc{\psib}{\overline{\psi}}

\begin{document}

\baselineskip 0.9cm

\cent{\large\bf ABOUT THE KINEMATICS OF SPINNING PARTICLES $^{(\dagger)}$}

\footnotetext{$^{(\dagger)}$ Work partially supported by INFN, MURST, CNR,
and by CNPq.}

\vs{0.5cm}

\begin{center}

{Giovanni SALESI

{\em Dipartimento di Fisica, Universit\`a  Statale di Catania,
95129--Catania, Italy; \ and

INFN--Sezione di Catania, Catania, Italy.}

\vs{0.2 cm}

and

\vs{0.2 cm}

Erasmo RECAMI

{\em Facolt\`a di Ingegneria, Universit\`a Statale di Bergamo,
24044--Dalmine (BG), Italy;

INFN--Sezione di Milano, Milan, Italy; \ and

Dept. of Applied Math., State University at Campinas, Campinas, S.P.,
Brazil.}}

\end{center}

\vs{3.0cm}

{\bf ABSTRACT \ --} \ Inserting the correct Lorentz factor into the
definition of the 4-velocity $v^\mu$ for spinning particles
entails new kinematical properties for $v^2$. The well-known
constraint (identically true for scalar particles, but entering also
the Dirac theory, and assumed {\em a priori} in all
spinning particle models)
  \ $p_{\mu}v^{\mu} = m$ \ is here derived in a self-consistent way.

\newpage

%
%
%

\section{The ``extended--like'' electron}

\

\h Since the works by Compton,$^{[1]}$  Uhlenbeck and Goudsmith,$^{[2]}$
Frenkel,$^{[2]}$ and Schr\"odinger$^{[3]}$
till the present times, many classical theories ---often quite different
among themselves from a physical
and formal viewpoint--- have been advanced for spinning particles.$^{\# 1}$
\footnotetext{$^{\# 1}$ Hereafter we shall often write
``electron'' or ``spinning particle'' instead  of the more
pertinent expression ``spin-$1 \over 2$ particle''.}

\h Following Bunge$^{[4]}$ , they can be divided into
three classes:

\h I) strictly {\em point-like} particle models

\h II) actual extended--type particle models
(``spheres'', ``tops'', ``gyroscopes'', and so on)

\h III)  mixed models for ``extended--like" particles, in which
the center of the {\em point-like} charge $\cal Q$ results to be spatially
distinct from the particle center-of-mass (CM).

\h Notice that in the theoretical approaches of type III ---which, being
in the middle
between classes I and II, could answer the dilemma posed by Barut at the top
of this paper---
the motion of $\cal Q$ does not coincide with the motion of the particle CM.
This peculiar feature has been actually found to be a characteristic for
the kinematics of spinning particles, and  is
known  as the zitterbewegung (zbw) motion.$^{[5-10]}$
The existence of such an internal motion is denounced,
besides by the presence itself of spin, by the
remarkable fact that in the ordinary Dirac theory the
particle four-impulse $p^\mu$ is {\em not} parallel to the four-velocity: \
$v^\mu \neq p^\mu/m$. \ Moreover, while \ $[{\imp},
{H}]=0$ \ so that $\imp$ is a conserved quantity,
$\vbf$ is {\em not} a constant of the motion: \ $[{\vbf}, {H}]\neq
0 \ \ ({\vbf} \equiv \albf \equiv \ga^0\gabf$ being the usual vector matrix
of the Dirac theory). \
Let us explicitly notice that, for the models belonging to class III,
assuming the zbw is equivalent$^{[7-9]}$ to splitting the motion
variables as follows (the dot meaning derivation with respect to the proper time
$\tau$):
\bb
x^\mu \equiv \xi^\mu + X^\mu \; ; \ \ \ \dot{x}^\mu \equiv v^\mu = w^\mu + V^\mu \ ,
\ee
where $\xi^\mu$ and $w^\mu\equiv \dot{\xi}^\mu$ describe
the {\em translational, external} or {\em drift} motion, i.e. the motion
of the CM,
whilst $X^\mu$ and $V^\mu \equiv \dot{X}^\mu$ describe the {\em internal} or
{\em spin} motion.\
From an electrodynamical point of view, the conserved electric current is
associated with the
trajectories of $\cal Q$ (i.e., to $x^\mu$), whilst the center of the
particle Coulomb field ---obtained$^{[10]}$  through a time average
over the field generated by the quickly oscillating charge---
is associated with the CM  (i.e., with $w^\mu$; and then, for free particles,
to the geometrical center of the helical trajectory).$^{\# 2}$  \
\footnotetext{$^{\# 2}$ From the classical--electrodynamics viewpoint, also in the
free case, the charge, moving along a non-rectilinear (helical) path, should
suffer a radiation-emission. Nevertheless, often  this is assumed {\em not}
to happen, in analogy
with the stationary atoms orbits, and because of the fact that no
external field
is responsible for the accelerations of $\cal Q$, whose motion is
``inertial'', in a way. \ However, it is also
possible to regard the charge as actually radiating, and at the same time
holding itself along stationary
``light-like'' orbits, because of a perfect balance
(when time-averaging on stochastic fluctuations)
between power emitted and power absorbed by any other accelerated
charge in the universe. Starting from these assumptions, once known
the numerical
value of the cosmological Hubble constant, in a recent work$^{[11]}$ the
value of the Planck constant has been deduced.}
In such a way, it is $\cal Q$ which follows {\em the (total) motion}, whilst
the CM follows the {\em mean motion} only.
It is important to remark that the classical extended--like electron
of type III
is quite {\em consistent} with the standard Dirac theory. In fact the above
decomposition for the total motion is the classical analogue of two
well-known
quantum-mechanical procedures, i.e., the so-called {\em Gordon decomposition}
of the Dirac current, and the (operatorial) {\em decomposition of the
Dirac position operator}
proposed by Schr\"odinger in his pioneering works.$^{[3]}$
We shall show these points below.

\h The well-known Gordon decomposition of the Dirac current reads$^{[13]}$
(hereafter we shall choose units such that numerically $c=1$):
\bb
\psib\ga^\mu\psi = \frac{1}{2m}\,[\psib\po^\mu\psi - (\po^\mu\psib)\psi]
- \frac{i}{m}\po_\nu\,(\psib S^{\mu\nu}\psi)\; ,
\ee
$\psib$ being the ``adjoint'' spinor of $\psi$; \ quantity
$\po^\mu \equiv i\pa^\mu$
the 4-dimensional impulse operator; and $S^{\mu\nu} \equiv
\frac{i}{4}\,(\ga^\mu\ga^\nu - \ga^\nu\ga^\mu)$
the spin-tensor operator.
 \ The ordinary interpretation of eq.(2) is in total analogy
with the decomposition given in eq.(1).
The first term in the r.h.s. results to be associated with
the translational motion of the CM ({\em scalar} part of the current,
corrisponding to
the traditional Klein--Gordon current).
The second term in the r.h.s. results, instead, directly connected with
the existence of spin, and describes the zbw motion.

\h In the abovequoted papers, Schr\"odinger started from the Heisenberg equation
for the time evolution of the acceleration operator in Dirac theory
\bb
\abf \equiv \frac{\drm\vbf}{\drm t} \ug {i\over\hbar}\,[H, \vbf] \ug
{2i \over \hbar} \, (H\vbf - \imp)\; ,
\ee
where $H$ is equal as usual to $\vbf$$\cdot$$\imp + \bt\,m$
($\vbf\equiv\albf$). \
Integrating once this operator equation over time, after some
algebra one can obtain:
\bb
\vbf \ug H^{-1}\imp - {i \over 2} \hbar \, H^{-1}\abf\; ,
\ee
and, integrating it a second time, one obtains just the spatial part of
the decomposition:
\bb
\xbf \equiv \xibf + \Xbf
\ee
where (still in the operator formalism) it is
\bb
\xibf = \rbf + H^{-1}\imp t \; ,
\ee
related to the CM motion, \ and
\bb
\Xbf = {i \over 2} \hbar \, \etbf H^{-1} \; ,  \qquad  (\etbf \equiv \vbf -
H^{-1}\imp) \; ,
\ee
related to the zbw motion.

\h Besides their consistency with the quantum theory, the type III models
do easily entail the existence of spin, zbw and intrinsec magnetic
moment for the electron,
while these properties are hardly predicted by making recourse to the
point-like--particle theories of class I.
 \ The ``extended--like" electron models of class III are at present
after fashion because of their possible generalizations to include
supersymmetry and superstrings.$^{[8]}$   Furthermore, the ``mixed'' models
seem to overcome the known non-locality problems involved by a relativistic
covariant picture for extended--type (in particular {\em rigid}) objects
of class II. \
Quite differently, the extended--like (class III) electron is
non-rigid and consequently variable in its ``shape ''and in its
characteristic ``size'', depending on the considered dynamical
situation. This is a priori consistent with  the appearance
in the literature of many different ``radii of the electron''.$^{\# 3}$
\footnotetext{$^{\# 3}$ In his book {\em The Enigmatic Electron},$^{[13]}$
M.H. McGregor lists at page 5 seven typical electron radii,
from the Compton radius to the ``classical'' and to the ``magnetic'' one.}
Because of all those reasons, therefore, the spinning particle to which we
shall refer in the next Section is described by a class III theory.\\

\

\section{New kinematical properties for the extended--like spinning particles}

\

\h We want now to analyze the formal and conceptual properties of a {\em new
definition} for the 4-velocity of our extended--like electron.
Such a new definition has been first adopted ---but without any emphasis---
in the papers by Barut et al.
dealing with a recent, important model for the relativistic classical
electron.$^{[7,8,9,10]}$  \ Let us consider the following.
At variance with the procedures followed in the literature
from Schr\"odinger's till our days, we have to make recourse {\em not to
the proper time of the
charge $\cal Q$, but rather to the proper time of the
center-of-mass}, i.e. to the time of the CM frame (CMF). \ $^{\# 4}$ \ \
\footnotetext{$^{\# 4}$ Let us recall that the CMF is the frame in which the
kinetic impulse vanishes identically, $\imp = 0$.  For spinning
particles, in general, it is {\em not} the ``rest'' frame, since the
velocity $\vbf$ is not necessarily zero in the CMF.}
As a consequence, if we examine the definition for the 4-velocity, quantity
$\tau$ in the
denominator of $v^\mu \equiv \drm x^\mu/ \drm \tau$ has to be
the latter proper time. \
Up to now ---with the exception of the above-mentioned papers by
Barut et al.--- in all
theoretical frameworks the Lorentz factor has been assumed to be equal to
$\sqrt{1 - {\bf v}^2}$, exactly as for the scalar particle case. \ On the
contrary, into the Lorentz factor it has to enter
${\bf w}^2$ instead of ${\bf v}^2$, \
quantity  ${\bf w} \equiv {\bf p}/{p^0}$ being the 3-velocity of the CM
with respect to the chosen frame [$p^0 \equiv {\cal E}$ is the energy]. \
By adopting the correct Lorentz factor, all the formulae containing
it {\em have to be rewritten, and they get a new physical meaning}. \
In particular, we shall show below that the new definition does actually
{\em imply}$^{\# 5}$ the following important
constraint, which ---holding identically for scalar particles--- is often
just {\em assumed} for spinning particles:
\footnotetext{$^{\# 5}$ For all plane wave solutions $\psi$ of the Dirac
equation, we have (labelling by $< >$ the corresponding
{\em local mean value} or {\em local density}): \
$p_{\mu}< \wide{v}^\mu > \equiv p_{\mu}\psi^{\dag} \wide{v}^\mu \psi \equiv
p_{\mu}\psi^{\dag} \ga^0 \ga^\mu \psi \equiv p_{\mu}\ov{\psi} \ga^\mu \psi
= m$.}
\bb
p_{\mu}v^{\mu} = m   \ ,
\ee
where $m$ is the {\em physical} rest mass of the particle (and not an
undefined mass-like quantity $M$.$^{\# 6}$\\
\footnotetext{$^{\# 6}$ Let make just an example, recalling that
Pa\v{v}si\v{c}$^{[8]}$ derived, from a lagrangian containing an
{\em extrinsic curvature}, the classical equation of the motion
for a rigid $n$-dimensional world-sheet in a curved background spacetime. \
Classical world-sheets describe {\em membranes} for
$n\geq3$, strings for $n=2$, and point particles for $n=1$. \
For the special case $n=1$, he found nothing but the traditional
Papapetrou equation for a classical spinning particle; also,
by ``quantization'' of the classical theory, he actually  derived
the Dirac equation.  In ref.[8], however, $M$ is not the observed
electron mass $m$:
and the relation between the two masses reads: $m= M+\mu H^2$, quantity
$\mu$ being the so-called {\em string rigidity}, while $H$ is the second
covariant derivative on the world-sheet.}

\h Our choice of the proper time $\tau$ may be supported by the
following considerations:\\

\h (i) The {\em light-like} zbw ---when the speed of $\cal Q$ is constant
and equal to
the speed of light in vacuum--- is certainly the preferred one
(among all the ``a priori'' possible internal motions) in the literature,
and to many authors it appears
the most adequate for a meaningful classical picture
of the electron.$^{\# 2}$ \ In some special theoretical
approaches,$^{[5,6,11,12]}$
the light speed is even regarded as the {\em quantum-mechanical} typical
speed for the zbw.
In fact, the Heisenberg principle in the relativistic domain$^{[14]}$
implies (not controllable) particle--antiparticle pair creations
when the (CMF) observation involves space distances of the order of a Compton
wavelenght. So that $\hbar / m$ is assumed to be the characteristic
``orbital" radius, and $2m / \hbar^2$ the (CMF) angular frequency,
of the zbw ---as first noticed by Schr\"odinger---; \ and the orbital motion of
$\cal Q$ is expected to be {\em light-like}.  \ \
Now, if the charge $\cal Q$ travels at the light speed, {\em its
proper time
of $\cal Q$ cannot exist}; while the proper time of
the CM (which travels at sub-luminal speeds) does exist. \
Adopting as time the proper time of $\cal Q$, as often done in the past
literature, automatically {\em excluded} a
ligh-like zbw. In our approach, by contrast, such zbw motions are
not excluded. \
Analogous considerations may hold for {\em Super-luminal} zbw speeds,
without any problem since the CM (which carries the
energy-impulse and the ``signal") is always endowed with a subluminal motion;

\h (ii) The indipendence between the center-of-charge and
the center-of-mass motion becomes evident by our definition. As a
consequence {\em the non-relativistic limit can be formulated by us in a
correct, and univocal, way}. Namely, by assuming the correct Lorentz
factor, one can immediately see
that the zitterbewegung can go on being a relativistic (in particular,
light-like) motion$^{\# 7}$
\footnotetext{$^{\# 7}$ This is perhaps connected with the
non-vanishing of spin in the non-relativistic limit, once we accept a
correlation between spin and zbw.}
even in the non-relativistic approximation: i.e., when $\imp \longr 0$.
 \ In fact, in the non-relativistic limit, we have to take
\[
\wbf^2 \ll 1 \ ,
\]
and {\em not} necessarily
\[
\vbf^2 \ll 1
\]
as usually assumed in the past literature;

\h (iii) Our proposed definition for the 4-velocity agrees with the natural
``classical limit" of the Dirac current. Actually, it has been used in those
models which (like Barut et al.') define velocity
{\em even at the classical level}
as the bilinear combination $\psib\ga^\mu\psi$,
via a direct introdution of {\em classical} spinors $\psi$. \
By the new definition, we shall be able to write the translational
term as $p^{\mu}/m$, with the {\em physical} mass in the denominator, exactly
as in the Gordon decomposition, eq.(2). \ Quite differently, in all the
theories adopting as time the proper time of $\cal Q$, it appears in the
denominator
the already mentioned {\em variable} mass $M$, which depends on the internal
zbw speed $V$ (see below);

\h (iv) The choice of the CM proper time constitutes a natural extension of
the ordinary procedure for relativistic scalar particles. In fact,
for spinless particles in relativity the 4-velocity is known to be univocally
defined as the derivative of 4-position with respect to the CMF
proper time (which is the only one available).\\

\h The most important reason in support of our definition turns out to be
the noticeable circumstance that {\em the old definition}
\bb
v_{\rm std}^\mu = (1/ \sqrt{1- \vbf^2}; \;\; \vbf / \sqrt{1- \vbf^2})\;
\ee
{\em seems to entail a mass varying with the internal zbw speed}.

\h But let us explicitate our new definition for $v^\mu$. \
The symbols which we are going to use possess the ordinary meaning; the
novelty is that now {\em the Lorentz factor $\drm\tau / \drm t$ will not be
equal to $\sqrt{1- \vbf^2}$, but instead to $\sqrt{1- \wbf^2}$ .} \
Thus we shall have:
\[
v^\mu\equiv \drm x^\mu/ \drm \tau \equiv (\drm t/ \drm \tau;\,
\drm \xbf / \drm \tau)
\equiv \left({\frac{\drm t}{\drm \tau}};\; \frac{\drm \xbf}{\drm t} \;
\frac{\drm t}{\drm \tau}\right)
\]
\bb
=(1/ \sqrt{1- \wbf^2}; \;\; \vbf / \sqrt{1- \wbf^2})\; . \;\;\;\;\;\; \quad
[\vbf \equiv \drm \xbf / \drm t]
\ee
For $w^\mu$ we can write:
\[
w^\mu\equiv \drm \xi^\mu/ \drm \tau \equiv (\drm t/ \drm \tau;\,
\drm \xibf / \drm \tau)
\equiv \left({\frac{\drm t}{\drm \tau}};\; \frac{\drm \xibf}{\drm t} \;
\frac{\drm t}{\drm \tau}\right)
\]
\bb
=(1/ \sqrt{1- \wbf^2}; \;\; \wbf / \sqrt{1- \wbf^2})\; ;\;\;\;\;\;\; \quad
[\wbf \equiv \drm \xibf / \drm t]
\ee
and for the 4-impulse:
\bb
p^\mu \equiv mw^\mu = m (1/ \sqrt{1- \wbf^2}; \;\; \wbf / \sqrt{1- \wbf^2})
 \; .
\ee
[In presence of an external field such relations remain valid,
provided that one
makes the ``minimal prescription": \ $p \longr p - eA$ \ (in the CMF we
shall have
$\imp - e\Abf = 0$ and consequently $\wbf = 0$, as above)].

\h Let us now examine the resulting impulse--velocity scalar product,
$p_\mu v^\mu$, which has to
be a Lorentz invariant, both with our $v$ and with the old $v_{\rm std}$. \
Quantity
$p \equiv (\vare;\; \imp)$ being the 4-impulse, and $M_1, M_2$ two relativistic
invariants, we may write:
\bb
p_\mu v^\mu \equiv M_1 \equiv \frac{\vare - \imp\cdot\vbf}{\sqrt{1- \wbf^2}}\; ,
\ee
or, alternatively,
\bb
p_\mu v^\mu_{\rm std} \equiv M_2 \equiv \frac{\vare - \imp\cdot\vbf}
{\sqrt{1- \vbf^2}}\; .
\ee
If we refer ourselves to the CMF, we shall have \ $\imp_{{\rm CMF}} =
\wbf_{{\rm CMF}} = 0$ \ (but $\vbf_{{\rm CMF}}
\equiv \Vbf_{{\rm CMF}} \neq 0$), \ and then
\bb
M_1 = \vare_{{\rm CMF}}
\ee
in the first case; and
\bb
\vare_{{\rm CMF}} = M_2\,\sqrt{1- \Vbf_{{\rm CMF}}^2}
\ee
in the second case. \ \ So,
we see that the invariant $M_1$ is actually a {\em constant}, and ---being
nothing but
the center-of-mass energy, $\vare_{{\rm CMF}}$--- it can be identified,
as we are going to prove, with
the physical mass $m$ of the particle. \ On the contrary, in the second
case (the standard one), the center-of-mass energy results to be
{\em variable} with the internal motion.

\h Now, from eq.(12) we have
\[
p_{\mu}v^{\mu} \equiv m w_{\mu}v^{\mu}
\]
and, because of eqs.(9-11),
\bb
p_{\mu}v^{\mu} \equiv m (1 - \wbf\cdot\vbf) / 1- \wbf^2 \ .
\ee
Since $\wbf$ is a vector component of the total
3-velocity $\vbf$, due to eq.(1), and $\wbf$  {\em is the orthogonal
projection} of $\vbf$ along the $\imp$-direction, we can write
\[
\wbf\cdot\vbf = \wbf^2 \ ,
\]
which, introduced into eq.(17), yields the constraint (8):
\[
p_{\mu}v^{\mu} = m \ .
\]

\h Quite differently, by use of the wrong Lorentz factor, we would
have got
\[
v^\mu = (1/ \sqrt{1- \vbf^2}; \;\; \vbf / \sqrt{1 - \vbf^2})
\]
and consequently
\[
p_{\mu}v^{\mu} \equiv m (1 - \wbf\cdot\vbf) / \sqrt{(1 - \wbf^2)(1 - \vbf^2)}
\]
\[
= m \sqrt{1 - \wbf^2} / \sqrt{1 - \vbf^2} \neq m \ .
\]

\h By recourse to the correct Lorentz factor, therefore, we succeeded in
showing that the important constraint \ $p_{\mu}v^{\mu} \ug m$, \
trivially valid for scalar particles, does hold for spinning particles too.
\\

\h Finally, we want to show that ---however--- the ordinary kinematical
properties of the Lorentz invariant $v_{\mu}v^{\mu}$ do not hold any longer
in the case of spinning particles, endowed with zitterbewegung. \ In fact,
it is easy to prove
that the ordinary constraint for scalar relativistic
particles ---$v^2$ constant in time and equal to $1$---
does {\em not} hold for spinning particles endowed with zbw.
\ Namely, if we choose as reference frame the CMF, in which  $\wbf = 0$,
we have [cf. definition (10)]:
\bb
v^\mu_{\rm CMF} \equiv (1; \Vbf_{{\rm CMF}}) \ ,
\ee
{\em wherefrom}, it being
\bb
v^2_{\rm CMF} \equiv 1 - \Vbf^2_{{\rm CMF}}\; ,
\ee
one can deduce the following {\em new constraints}:
\[
0 < v^2_{{\rm CMF}} (\tau) < 1 \;\;\; \Lef \;\;\; 0 < \Vbf_{{\rm CMF}}^2
(\tau) < 1 \;\;\;\;\;\; \quad\mbox{(``time-like")}
\]
\bb
v^2_{{\rm CMF}}(\tau)=0 \quad \Lef \quad \Vbf^2_{{\rm CMF}} (\tau) = 1
\;\;\;\;\; \ \ \ \ \ \ \ \;
\quad\mbox{ \ \ (``light-like")}
\ee
\[
v^2_{{\rm CMF}} (\tau) < 0 \quad \Lef \quad \Vbf^2_{{\rm CMF}} (\tau) > 1 \ .
\;\;\;\;\; \ \ \ \ \ \ \ \ \quad\mbox{ \ \ (``space-like")}
\]
Notice that, since the square of the total 4-velocity is invariant and in
particular it is $v^2_{{\rm CMF}} =
v^2$, these new constraints for $v^2$ will be valid in any frame:
\[
0 < v^2 (\tau) < 1 \;\;\;\;\;\;\;\;\;\;\;\;
\quad\mbox{(``time-like'')}
\]
\bb
\ \ v^2 (\tau)=0  \ \ \;\;\;\;\;\;\;\;\;\;\;\;\;
\quad\mbox{(``light-like'')}
\ee
\[
\ \ \ \ v^2 (\tau) < 0 \ .  \;\;\;\;\;\;\;\;\;\;\;\;\;
\quad\mbox{(``space-like'')}
\]
Let us examine the manifestation and consequences of such new constraints
in a specific example: namely, in the
already mentioned theoretical model by Barut--Zanghi$^{[7]}$
which did implicitly adopt as time the proper time of the CMF. \
In this case we get that in general it is $v^2\neq 1$. \ And in fact
one obtains the important relation:$^{[9]}$
\bb
v^2 = 1 - \frac{\ddot{v}_\mu v^\mu}{4 m^2} \ .
\ee
In particular,$^{[10]}$ in the light-like case it is \
$\ddot{v}_\mu v^\mu = 4 m^2$ \ and therefore \ $v^2 = 0$.

\h Going back to eq.(19), notice that now quantity $v^2$ is no longer related
now to the {\em external} CM speed $|\wbf|$, but on the contrary to
the {\em internal} zitterbewegung speed $|\Vbf_{{\rm CMF}}|$. \ Notice at last
that, in general ---and at variance with the scalar case---
the value of $v^2$ {\em is not constant in time} any longer, but varies
with $\tau$
(except when ${\Vbf^2_{{\rm CMF}}}$ is constant in time).

\vs{2.5 cm}

{\bf Acknowledgements}\\

This paper is dedicated to the memory of Asim O. Barut. \ The authors
acknowledge stimulating discussions with E.C. de Oliveira,
M. Pav\v{s}i\v{c}, F. Raciti, D. Wisnivesky and particularly,
J. Keller, W.A. Rodrigues and J. Vaz.
\ For the kind cooperation, thanks are also due to G. Andronico, I. Arag\'on,
M. Baldo, A. Bonasera, M. Borrometi, A. Bugini, F. Catara, L. D'Amico,
G. Dimartino, C. Dipietro,  M. Di Toro, P. Falsaperla, G. Giuffrida,
L. Lo Monaco, R.L. Monaco, Z. Oziewicz, M. Pignanelli, E. Previtali,
G.M. Prosperi, M. Sambataro, S. Sambataro,  M. Scivoletto, R. Sgarlata,
G. Tagliaferri, E. Tonti, R. Turrisi and M.T. Vasconselos.

\end{document}